# Charge density wave modulated third-order nonlinear Hall effect in 1$T$-VSe$_2$ nanosheets

Zhao-Hui Chen,[1,][*] Xin Liao,[1,][*] Jing-Wei Dong,[1] Xing-Yu Liu,[1] Tong-Yang Zhao,[1] Dong Li,[1,2]
An-Qi Wang,[1,][†] and Zhi-Min Liao[1,3,][‡]

[1]*State Key Laboratory for Mesoscopic Physics and Frontiers Science Center for Nano-optoelectronics,
School of Physics, Peking University, Beijing 100871, China*
[2]*Academy for Advanced Interdisciplinary Studies, Peking University, Beijing 100871, China*
[3]*Hefei National Laboratory, Hefei 230088, China*

We report the observation of a pronounced third-order nonlinear Hall effect (NLHE) in 1$T$-phase VSe$_2$ nanosheets, synthesized using chemical vapor deposition (CVD). The nanosheets exhibit a charge density wave (CDW) transition at ∼77 K. Detailed angle-resolved and temperature-dependent measurements reveal a strong cubic relationship between the third-harmonic Hall voltage $V_\perp^{3\omega}$ and the bias current $I^\omega$, persisting up to room temperature. Notably, the third-order NLHE demonstrates a twofold angular dependence and significant enhancement below the CDW transition temperature, indicative of threefold symmetry breaking in the CDW phase. Scaling analysis suggests that the intrinsic contribution from the Berry connection polarizability tensor is substantially increased in the CDW phase, while extrinsic effects dominate at higher temperatures. Our findings highlight the critical role of CDW-induced symmetry breaking in modulating quantum geometric properties and nonlinear transport phenomena in VSe$_2$, paving the way for future explorations in low-dimensional quantum materials.

## I. INTRODUCTION

Two-dimensional (2D) transition metal dichalcogenides (TMDs) have garnered significant attention due to their unique electronic, optical, and catalytic properties [1–5]. These include the much-studied charge density wave (CDW) state [6] which is a collective physical phenomenon and manifests as periodic modulations of electronic charge density within a material. The electronic charge density modulation is ubiquitously accompanied by the spatial lattice distortion and the symmetry breaking of a system. The CDW-induced symmetry breaking could give rise to emerging orders and fascinating phenomena. For instance, the CDW phase exhibits intimate correlation with the ferromagnetism [7,8] and superconductivity [9–11] in certain materials, and plays a significant role in the three-dimensional (3D) quantum Hall effect of ZrTe$_5$ [12,13].

Recent advances have revealed the significance of the nonlinear Hall effect (NLHE) [14–16] in TMDs, where reduced dimensionality and enhanced electron correlations can lead to exotic quantum geometry properties. While the conventional Hall effect is linear and arises due to the Lorentz force acting on charge carriers in the presence of a magnetic field, the NLHE can occur in nonmagnetic materials without an external magnetic field, originating from the intrinsic properties of the material's band structure, such as Berry curvature dipole [17–22] and Berry connection polarizability [16,23,24]. In this context, 1$T$-vanadium diselenide (VSe$_2$) stands out as an intriguing candidate. 1$T$-VSe$_2$, a member of the TMD family, undergoes a CDW transition with decreasing temperature [25–30], which significantly alters its electronic structure and symmetry properties [31–38]. This CDW phase transition provides an excellent platform to explore the interplay between symmetry breaking, band geometric properties, and nonlinear transport phenomena.

Here, we reveal the CDW-induced symmetry breaking through the third-order NLHE measurements in 1$T$-phase VSe$_2$ nanosheets synthesized by chemical vapor deposition (CVD). Our comprehensive characterization confirms the high-quality crystalline nature of the nanosheets and the presence of the CDW transition. Through systematic angle-resolved and temperature-dependent measurements, we demonstrate that the third-order NLHE exhibits a strong cubic dependence on the bias current and a twofold angular dependence. Notably, the effect is significantly enhanced below the CDW transition temperature. The enhanced NLHE, along with its twofold angular dependence, provides evidence for the breaking of the threefold crystal symmetry following the CDW transition in 1$T$-VSe$_2$.

## II. EXPERIMENTAL METHOD

*Crystal growth and characterization.* The 1$T$-VSe$_2$ crystal has a hexagonal layered structure and belongs to the trigonal space group $P$-3$m$1. The Se-V-Se monolayers are arranged in an AA-stacking order along the $c$ axis, with each V atom surrounded by six Se atoms in the individual layer [Fig. 1(a)]. Single-crystal 1$T$-VSe$_2$ nanosheets were synthesized using the CVD technique (see Appendix A). The obtained nanosheet thickness ranges from several nanometers to several tens of nanometers. The nanosheets were then characterized by scanning electron microscopy (SEM) with *in situ* energy-dispersive X-ray spectroscopy (EDS). As shown in Figs. 1(b) and 1(c), the SEM image of a typical VSe$_2$ nanosheet features a half-hexagonal structure, while the EDS elemental analysis indicates an atomic ratio of Se : V ≈ 2 : 1, consistent with the chemical composition of VSe$_2$. The high-resolution transmission electron microscopy (TEM) image shows the interplanar spacings of 1.69 and 3.0 Å aligning with the (110) and (100) planes of VSe$_2$, respectively [39]. The corresponding selected-area electron diffraction (SAED) pattern reflects the single-crystal nature of VSe$_2$ with (001) surface orientation. The Raman spectrum [Fig. 1(e)] exhibits two prominent peaks at 207 and 139 cm$^{-1}$, corresponding to the out-of-plane $A_{1g}$ mode and the in-plane $E_g$ mode of few-layer 1$T$-VSe$_2$, respectively [40,41]. The additional peak at 257 cm$^{-1}$ arises from the oxidation of VSe$_2$ under ambient condition, a common observation in previous works [40,42].

---

[*]These authors contributed equally to this work.
[†]Contact author: anqi0112@pku.edu.cn
[‡]Contact author: liaozm@pku.edu.cn

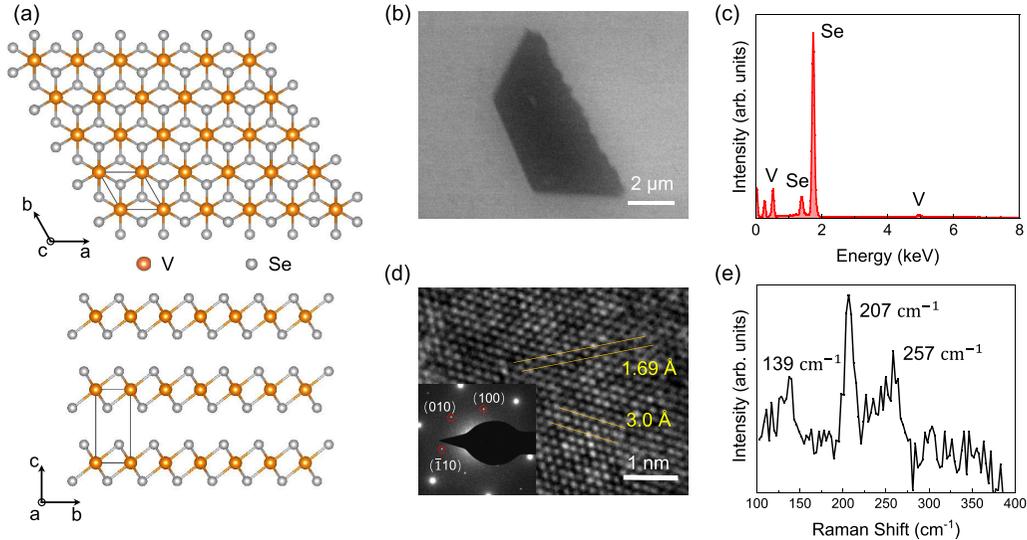

FIG. 1. (a) Top (upper panel) and side (lower panel) views of the atomic structure of 1$T$-VSe$_2$. (b) SEM image of 1$T$-VSe$_2$ nanosheet. (c) EDS spectrum of the VSe$_2$ nanosheet. (d) High-resolution TEM image of the VSe$_2$ nanosheet. Inset: The corresponding SAED pattern indicates that the naturally grown surface is the (001) plane. (e) Raman spectrum of a 12 nm-thick VSe$_2$ nanosheet.

*Device fabrication and transport measurements.* Individual VSe$_2$ nanosheet was first transferred onto a Si/SiO$_2$ substrate. The circular disc electrodes were patterned using electron beam lithography, and 5/60 nm Ti/Au layers were then evaporated on a VSe$_2$ nanosheet after *in situ* Ar$^+$ etching to remove the oxidation and establish Ohmic contact. Nanosheet devices were measured in an Oxford cryostat with a variable temperature insert and a superconducting magnet. Stanford Research Systems SR830 and SR865A lock-in amplifiers were used to measure the first-, second-, and third-harmonic voltage signals. The driving frequency of the bias current is fixed at 17.777 Hz, unless otherwise specified.

### III. RESULTS AND DISCUSSION

We first investigated NLHE in the disc device shown in Fig. 2(a). The thickness of the 1$T$-VSe$_2$ nanosheet was confirmed by atomic force microscopy (AFM) to be approximately 9 nm. As depicted in Fig. 2(b), an alternating current $I^\omega$ was applied between two opposite electrodes (S and D), while the Hall voltage was measured between another pair of transverse electrodes (A and B). Figure 2(c) shows the temperature dependence of longitudinal resistance in this device. The resistance $R_\parallel$ exhibits an overall metallic behavior with a clear hump feature due to the CDW transition at $T_p \sim$ 77 K (indicated by the red arrow), quantitatively consistent with previous studies [43]. The CDW transition temperature $T_p$ is identified as the temperature at which the differentiate $dR_\parallel/dT$ achieves a minimum value [Fig. 2(d)]. The residual resistance ratio (RRR = $R_{300\,\mathrm{K}}/R_{1.6\,\mathrm{K}}$) is about 3.67, smaller than that reported for bulk VSe$_2$ [43–45], due to enhanced disorder effects in the thin nanosheet during the microfabrication process [43]. Figures 2(e) and 2(f) present the current-voltage characteristics from the fundamental to third-harmonic frequencies measured at 1.6 K. At the fundamental frequency, the longitudinal voltage $V_\parallel$ exhibits a linear relation with the bias current $I^\omega$ [Fig. 2(e)]. The observed finite $V_\perp$ should originate from the misalignment of the current direction with the crystal axes and intrinsic resistance anisotropy of 1$T$-VSe$_2$. As extending to high-order Hall signals, a pronounced third-order transverse voltage $V_\perp^{3\omega}$ is observed which scales cubically with $I^\omega$ and dominates over the second-order transverse voltage $V_\perp^{2\omega}$. The frequency independence of $V_\perp^{3\omega}$ rules out the capacitive coupling effect (see Appendix B).

To further investigate the characteristics of the third-order NLHE, angle-resolved measurements at 1.6 K were conducted by rotating the measurement framework while maintaining the relative position of electrodes unchanged (S-A-D-B in clockwise sequence), thereby recording signals for different lattice directions. The current is injected to the nanosheet at angle $\theta$ [inset of Fig. 2(a)]. The voltage $V_\perp^{3\omega}$ shows a cubic relation with the bias current $I^\omega$ for each current direction [Fig. 3(a)]. Figure 3(b) presents the linear dependence of $V_\perp^{3\omega}$ on $V_\parallel^3$, where the slope varies between different directions. To clearly reveal the angle dependence of third-order nonlinear Hall response, we plot the slope $V_\perp^{3\omega}/V_\parallel^3$ as a function of angle $\theta$ in Fig. 3(e). It is found that $V_\perp^{3\omega}/V_\parallel^3$ possesses a twofold angular dependence, similar to the case of first-harmonic longitudinal and transverse resistances, i.e., $R_\parallel(\theta)$ and $R_\perp(\theta)$ in Figs. 3(c) and 3(d). The observed twofold angular dependence is not consistent with trigonal lattice structure, suggesting the threefold symmetry breaking at low temperature. This is expected by considering that the emerging CDW in 1$T$-VSe$_2$ is not identical along three wave vector directions. The anisotropic CDW modulation would break the original rotational invariance of the crystal lattice and result in lower-symmetry structure of the charge distribution [31,32,46,47]. The material might conduct electricity more easily along the direction of the CDW than perpendicular to it. It is found that the $R_\parallel(\theta)$ achieves a minimum near $\theta = 0°$ in this nanosheet, which may correspond to the preferential direction of the CDW in the plane. By contrast, $\theta = 90°$ is perpendicular to the CDW direction and leads to the least conductance (maximum $R_\parallel$). In the two directions, the $R_\perp(\theta)$ and $V_\perp^{3\omega}/V_\parallel^3$ almost vanish, similar to the case of materials with a $Pm$ space group such as WTe$_2$ [23,48], MoTe$_2$ [23], and TaIrTe$_4$ [49], further confirming the trigonal symmetry breaking in the CDW phase.

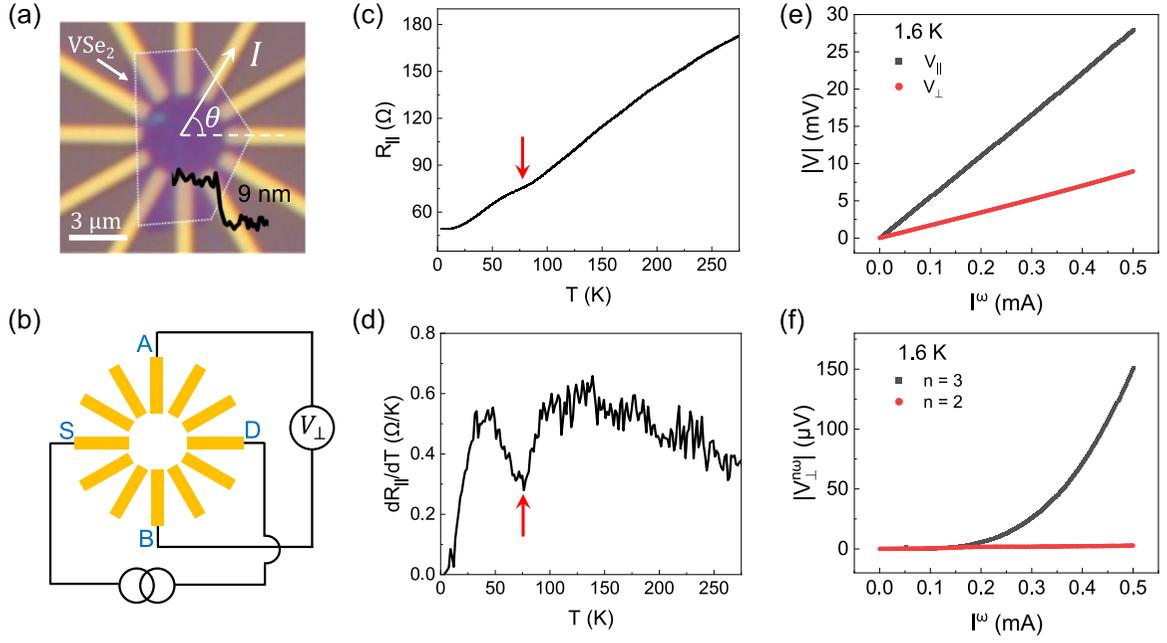

FIG. 2. (a) Optical image of the studied VSe$_2$ nanosheet device, where the nanosheet is outlined by a white dashed box. The electric current is applied into the nanosheet at angle $\theta$ as indicated. Inset shows the thickness profile indicated by a black line. (b) An alternating current $I^\omega$ with driving frequency $\omega$ is applied between the S and D electrodes while measuring the transverse voltage between the A and B electrodes. (c) Temperature-dependent longitudinal resistance of VSe$_2$ nanosheet. (d) The first derivative of resistance $dR_\parallel/dT$. (e) The first-harmonic longitudinal voltage $V_\parallel$ and transverse voltage $V_\perp$ as a function of bias current $I^\omega$ at 1.6 K. (f) Second- (red curve) and third-harmonic (black curve) $V_\perp^{n\omega}$ vs $I^\omega$ at 1.6 K. Current is applied along $\theta = 60°$ in (c)–(f).

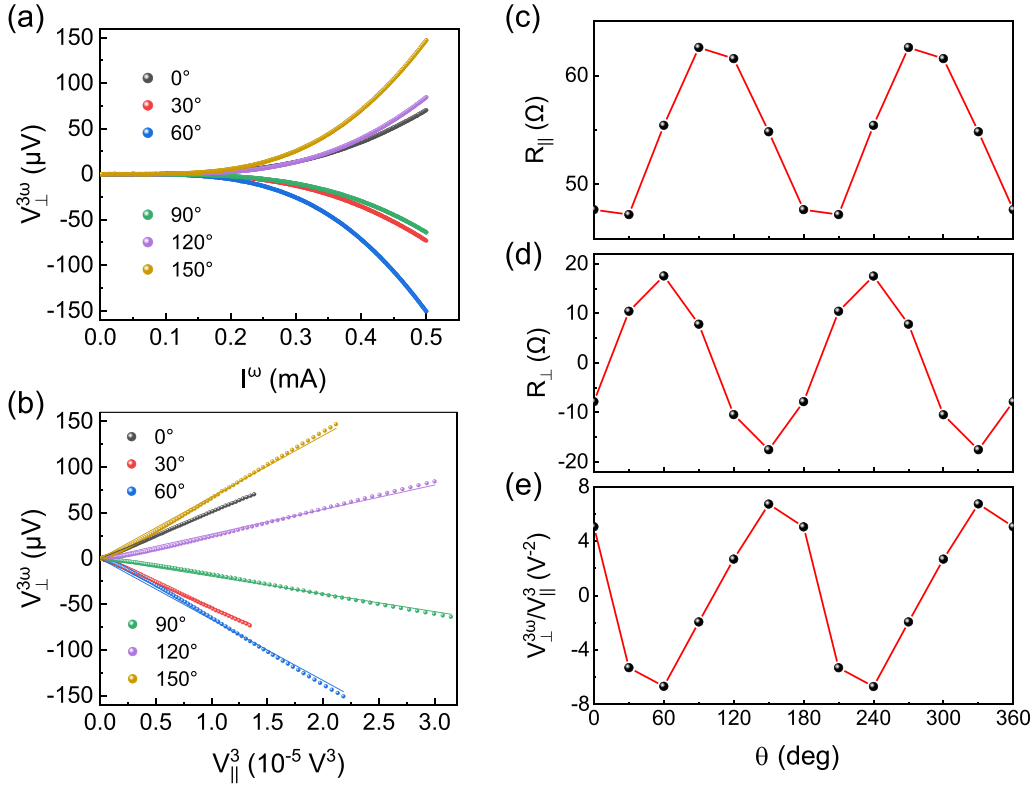

FIG. 3. (a) The angle-dependent third-harmonic Hall voltage $V_\perp^{3\omega}$. (b) $V_\perp^{3\omega}$ depends linearly on the cube of the first-harmonic $V_\parallel$ for different angles. (c)–(e) Angular dependence of the (c) longitudinal resistance $R_\parallel$, (d) transverse resistance $R_\perp$, and (e) third-harmonic Hall effect $V_\perp^{3\omega}/V_\parallel^3$, respectively. To obtain the angular dependence of resistance and NLHE in the disc device, the longitudinal voltage $V_\parallel$ for a given $\theta$ was first symmetrized between two opposing current directions, that is $[V_\parallel(\theta) + V_\parallel(\theta + 180°)]/2$, to eliminate the effects of longitudinal-transverse coupling. The experimental data were obtained at 1.6 K.

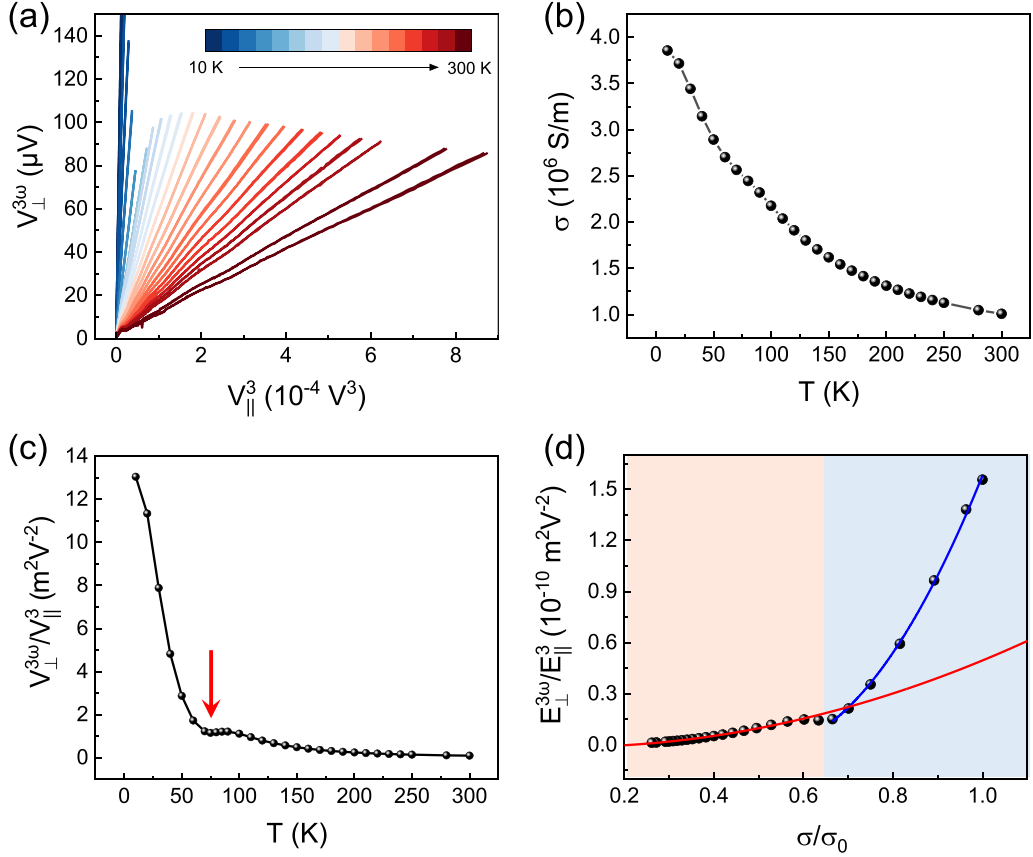

FIG. 4. (a) Temperature dependence of the third-harmonic NLHE. (b) The conductivity $\sigma$ as a function of $T$. (c) The slope of $V_\perp^{3\omega}/V_\parallel^3$ as a function of $T$. A kink (indicated by the red arrow) is clearly observed near 75 K. (d) $E_\perp^{3\omega}/E_\parallel^3$ vs $\sigma/\sigma_0$ in the VSe$_2$ nanosheet. Black curves just connect the experimental data points. Red (blue) curve represents the parabolic fitting of experimental data in the high-temperature (low-temperature) regime. The high-temperature (light red) and low-temperature (light blue) regions are divided by the occurrence of the CDW transition at $T_p$. Current is applied along $\theta = 0°$ in (a)–(d).

To reveal the underlying mechanism of the observed third-order NLHE, temperature-dependence measurements and scaling law analysis were performed. It is found that the third-harmonic Hall voltage $V_\perp^{3\omega}$ can sustain up to room temperature [Fig. 4(a)]. Besides, the $V_\perp^{3\omega}$ exhibits a linear relation with $V_\parallel^3$ under all temperatures. The longitudinal conductivity [Fig. 4(b)], calculated by $\sigma = L/(R_\parallel S)$ (where $L$ and $S$ are the length and cross-sectional area of the channel, respectively), decreases with increasing the temperature due to the increased carrier scatterings. The slope of $V_\perp^{3\omega}$ versus $V_\parallel^3$ exhibits a rapid decay below 75 K, which is roughly the CDW transition temperature $T_p$, while decreasing slowly above 75 K [Fig. 4(c)]. Figure 4(d) presents the evolution of $E_\perp^{3\omega}/E_\parallel^3$ against $\sigma/\sigma_0$, where $E_\perp^{3\omega} = V_\perp^{3\omega}/L$, $E_\parallel = V_\parallel/L$, and $\sigma = L/(R_\parallel S)$ with $\sigma_0$ the zero-temperature conductivity. According to previous works [24,50,51], the third-order NLHE $E_\perp^{3\omega}/E_\parallel^3$ is expected to show a parabolic dependence on $\sigma/\sigma_0$. Specifically, the parabolic relation is formulated as $\frac{E_\perp^{3\omega}}{E_\parallel^3} = A_0 + A_1(\frac{\sigma}{\sigma_0}) + A_2(\frac{\sigma}{\sigma_0})^2$, where $A_0$ includes the intrinsic contribution from the Berry connection polarizability tensor and associates with the field-induced Berry curvature dipole; $A_1$ and $A_2$ are linked with extrinsic contributions like impurity skew scattering (see Appendix C). The parabolic dependence can well fit the data above $T_p$ [red curve in Fig. 4(d)] but fails in the case below $T_p$, which exhibits a much larger value of $E_\perp^{3\omega}/E_\parallel^3$ than expected.

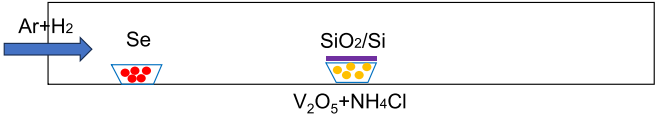

FIG. 5. Schematic diagram of the CVD growth of VSe$_2$.

The enhanced third-order NLHE may stem from the threefold symmetry breaking induced by the CDW phase, which has been observed in angular dependence measurements [Figs. 3(c)–3(e)]. This lower symmetry can enhance the third-order NLHE, since the element of susceptibility tensor $\chi$, which describes the relationship between the applied electric fields and the induced nonlinear Hall currents, can have nonzero components that were zero in a higher-symmetry phase. The low-temperature range can be fitted by another set of $A_i$ [blue curve in Fig. 4(d)], indicating the distinct underlying mechanism before and after CDW transition. The significantly increased $|A_0|$ below $T_p$ hints at the increased contribution of intrinsic components (see Appendix C) such as the Berry connection polarization (BCP) effect, in the CDW phase, while above $T_p$, the obtained three scaling parameters $|A_2| \gg |A_1| \sim |A_0|$, with an order of magnitude difference, indicate the dominance of the non-Gaussian skew scattering [24,50,51]. Such behavior sheds light on the variation of quantum geometry arising from the CDW transition-induced symmetry breaking and warrants further investigation in future studies.

## IV. CONCLUSIONS

In summary, we have demonstrated a significant third-order NLHE in 1$T$-VSe$_2$ nanosheets, evidenced by a strong cubic relationship between the third-harmonic Hall voltage $V_\perp^{3\omega}$ and the bias current $I^\omega$, which persists up to room temperature. Angle-resolved measurements indicate a marked enhancement and twofold angular dependence of the third-order NLHE below the CDW transition temperature of ∼77 K, suggesting a crucial role of CDW-induced symmetry breaking. Temperature-dependent measurements and scaling law analysis reveal that the intrinsic contribution of the Berry connection polarizability tensor significantly increases in the CDW phase, while extrinsic mechanisms dominate at higher temperatures. Our findings underscore the critical impact of CDW-induced symmetry breaking on nonlinear transport properties and highlight the potential of VSe$_2$ nanosheets in future electronic devices leveraging higher-order nonlinear responses.

## ACKNOWLEDGMENTS

This work was supported by the National Natural Science Foundation of China (Grants No. 62425401, No. 62321004, and No. 12204016) and Innovation Program for Quantum Science and Technology (Grant No. 2021ZD0302403).

## APPENDIX A: CRYSTAL GROWTH VIA THE CVD TECHNIQUE

1$T$-VSe$_2$ nanosheets were grown using the chemical vapor deposition (CVD) method. The synthetic sources were selenium powder and vanadium oxide powder (V$_2$O$_5$). The VSe$_2$ nanosheets were synthesized in a single-zone horizontal tube furnace equipped with a 1-in. quartz tube as shown in Fig. 5. The vanadium oxide powder (V$_2$O$_5$, 0.5 mg) with the sublimed-salt (NH$_4$Cl, 10 mg) in a quartz boat was placed at the center of the furnace. A clean SiO$_2$/Si substrate is placed on the boat facedown above the vanadium precursor. Selenium powder in another quartz boat was placed upstream of the furnace at a distance of 9.5 cm from the furnace center. Before heating, the air in the quartz tube was flushed with a mixture of H$_2$/Ar (2%) for 10 min. The furnace was heated to 440 °C in 10 min with H$_2$/Ar carrier gas at a flow rate of 100 SCCM (cubic centimeter per minute at STP), and kept for 20 min. After growth, the furnace was naturally cooled down to room temperature.

## APPENDIX B: FREQUENCY AND THICKNESS DEPENDENCES OF THIRD-ORDER NLHE

The third-order NLHE in the 1$T$-VSe$_2$ device was measured under alternating currents with different frequencies (17.777–177.777 Hz), as shown by Fig. 6(a). The $V_\perp^{3\omega}$ is found independent of driving frequency which can safely rule out the capacitive coupling effect. Moreover, we find that the third-order NHE is significantly stronger in thin nanosheets compared to thick ones [Fig. 6(b)].

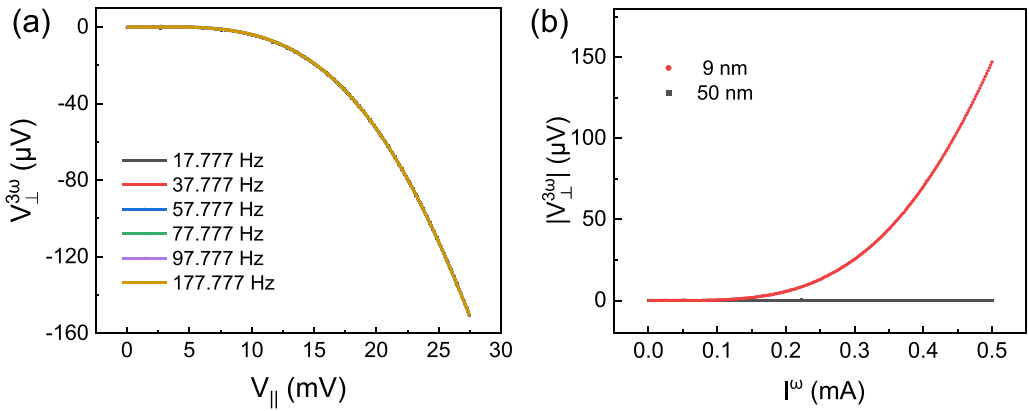

FIG. 6. (a) The third-harmonic Hall voltage $V_\perp^{3\omega}$ vs $V_\parallel$ under different driving frequencies. (b) The third-order NLHE in 9 nm-thick and 50 nm-thick VSe$_2$ nanosheets.

TABLE I. Scaling parameters obtained from parabolic fittings in Fig. 4(d).

| Temperature region | $A_0$ ($10^{-11}$ m²V⁻²) | $A_1$ ($10^{-11}$ m²V⁻²) | $A_2$ ($10^{-11}$ m²V⁻²) |
|---|---|---|---|
| Below $T_p$ | 16.7 | –67.0 | 66.1 |
| Above $T_p$ | –0.1 | –0.7 | 5.8 |

## APPENDIX C: SCALING LAW ANALYSIS ON THE OBSERVED NLHE

The scaling relation between $E_\perp^{3\omega}/E_\parallel^3$ and $\sigma/\sigma_0$ can be expressed as $\frac{E_\perp^{3\omega}}{E_\parallel^3} = A_0 + A_1(\frac{\sigma}{\sigma_0}) + A_2(\frac{\sigma}{\sigma_0})^2$, where $A_i$ ($i = 0, 1, 2$) are scaling parameters and $\sigma_0$ is the zero-temperature conductivity. Specifically, the three coefficients are given by $A_0 = C_{\text{int}} + C_1^{\text{sj}} + C_{11}^{\text{sk},1}$, $A_1 = C_{01}^{\text{sk},1} - 2C_{11}^{\text{sk},1} + C_0^{\text{sj}} - C_1^{\text{sj}}$, and $A_2 = C^{\text{sk},2}\sigma_0 + C_{00}^{\text{sk},1} + C_{11}^{\text{sk},1} - C_{01}^{\text{sk},1}$, here $C_{\text{int}}$ denotes the intrinsic contribution associated with BCP, $C_i^{\text{sj}}$ the extrinsic contribution from the side-jump mechanism, and $C_{ij}^{\text{sk},1}$ and $C^{\text{sk},2}$ from the Gaussian and non-Gaussian skew scattering, respectively. The indices $i$, $j$ taking values of 0 or 1 denote the impurity or phonon scattering source. For the VSe$_2$ nanosheet studied here, its high crystal quality indicates the weak side-jump effects, which are usually prominent in dirty metals. Therefore, we can exclude the side jump as the major factor influencing the three scaling parameters. Then the scaling parameters are simplified as $A_0 = C_{\text{int}} + C_{11}^{\text{sk},1}$, $A_1 = C_{01}^{\text{sk},1} - 2C_{11}^{\text{sk},1}$, and $A_2 = C^{\text{sk},2}\sigma_0 + C_{00}^{\text{sk},1} + C_{11}^{\text{sk},1} - C_{01}^{\text{sk},1}$.

As shown in Fig. 4(d), the parabolic scaling formula can independently fit the experimental data of $E_\perp^{3\omega}/E_\parallel^3$ versus $\sigma/\sigma_0$ below and above $T_p$. The extracted scaling parameters $A_i$ are listed in Table I. It is found that $|A_0|$ is two orders of magnitude larger at low temperatures (below $T_p$) compared to high temperatures (above $T_p$). Considering the reduced phonon scattering ($C_{11}^{\text{sk},1}$) at low temperatures, the enhancement of $|A_0|$ may arise from the significantly increased intrinsic BCP contribution ($C_{\text{int}}$) below $T_p$. As for the high-temperature range (above $T_p$), the three scaling parameters exhibit $|A_2| \gg |A_1| \sim |A_0|$ with an order of magnitude difference. The large value of $|A_2|$ beyond $|A_1|$ and $|A_0|$ hints at the dominance of the non-Gaussian skew scattering, which only takes place in $A_2$ and would lead to $A_2$ far exceeding the $A_0$ and $A_1$.